\documentclass{ws-ijmpa}
\renewcommand{\d}{\partial}

\newcommand{\pim}{\pi_{\scriptscriptstyle{-}}}
\newcommand{\pin}{\pi_{\scriptscriptstyle{0}}}
\newcommand{\Kp}{K_{\scriptscriptstyle{+}}}

\newcommand{\Kn}{K_{\scriptscriptstyle{0}}}
\newcommand{\Mpic}{M_{\pi^{\scriptscriptstyle{+}}}}
\newcommand{\Mpin}{M_{\pi^{\scriptscriptstyle{0}}}}
\newcommand{\MKc}{M_{\scriptscriptstyle{K^+}}}
\newcommand{\MKn}{M_{\scriptscriptstyle{K^0}}}
\newcommand{\Lagr}{\mathcal{L}}
\newcommand{\Order}[1]{\mathcal{O}#1}
\newcommand{\order}[1]{{\it o}#1}
\newcommand{\nn}{\nonumber}
\newcommand{\sss}{\scriptscriptstyle}
\begin{document}

\markboth{Julia Schweizer}
{Spectrum and Decays of Hadronic Atoms}

%
\catchline{}{}{}{}{}
%

\title{SPECTRUM AND DECAYS OF HADRONIC ATOMS}

\author{\footnotesize JULIA SCHWEIZER\footnote{
Present address: Institute for Theoretical Physics, University of  Vienna, Boltzmanngasse 5, A-1090 Vienna, Austria}}

\address{
Institute for Theoretical Physics, University of Bern, Sidlerstr. 5\\
CH-3012 Bern, Switzerland}

\maketitle


\begin{abstract}We describe the spectra and decays of $\pi^+\pi^-$ and $\pi^\pm K^\mp$ atoms
within a non-relativistic effective field theory. The evaluations of the
energy shifts
and widths are performed at next-to-leading order in isospin symmetry
breaking. The prediction for the lifetime of the $\pi^\pm K^\mp$ atom in its
ground-state yields $\tau=(3.7\pm0.4)\cdot10^{-15}$s.
\end{abstract}

\section{Introduction}
Hadronic atoms are highly non-relativistic loosely
bound systems of hadrons mainly formed by the Coulomb interaction. The study
of these hadronic bound states has seen growing interest during recent years,
because they provide important information on the behavior of QCD at low-energy. The DIRAC collaboration
\cite{Adeva:1994xz} at CERN measures the lifetime of the $\pi^+\pi^-$ atom,
and plans to determine the $\pi\pi$ scattering lengths difference $|a_0^0-a_0^2|$ at $5\%$
accuracy. Particularly exiting is the fact that in this manner the current
understanding of the SU(2)$\times$SU(2) chiral symmetry breaking in QCD can be
verified experimentally. Another experiment on hadronic atoms is presently
performed at PSI. The pionic hydrogen collaboration \cite{PSI} plans to
measure the strong energy shift and decay width of the $\pi^-p$ atom in its ground-state at the $1\%$ level. This high precision
measurement aims to determine the S-wave $\pi N$ scattering lengths.
Finally, the DEAR collaboration \cite{Bianco:1998wb} at the DA$\Phi$NE
facility measures the strong energy shift and width of the ground-state in
kaonic hydrogen. Here, the goal is to
extract the $K N$ S-wave scattering lengths.

In order to extract the scattering lengths from such precision
measurements, the theoretical expressions for the spectrum and decays must be known to a
precision that matches the experimental accuracy. Nearly fifty years ago, Deser {\it et al.} \cite{Deser:1954vq} derived the
leading order formulae for the decay width and the energy shift in pionic
hydrogen. Similar relations exist for $\pi^+\pi^-$ and $\pi^- K^+$ atoms
\cite{Palfrey:kt,Bilenky:zd}, which decay predominantly through the strong
transitions $\pi^+\pi^-\rightarrow\pi^0\pi^0$ and $\pi^-K^+\rightarrow\pi^0
K^0$, respectively. Theoretical investigations on the
spectrum and the decay of pionium have been performed beyond leading order
in several settings. In particular, the
lifetime of pionium was studied by the use of the Bethe-Salpeter equation
\cite{Lyubovitskij:1996mb} and in the framework of the
quasipotential-constraint theory approach \cite{Jallouli:1997ux}. The width of the $\pi^+\pi^-$ atom has also been analyzed within a
non-relativistic effective field theory
\cite{Labelle:1998gh,Eiras:2000rh,Gasser:2001un}, which was originally
developed for bound states in general by Caswell and Lepage
\cite{Caswell:1985ui}. The average momenta of the pions circulating in
pionium are very small, of the order of $\alpha \Mpic$, where $\alpha$ denotes
the fine-structure constant. For this reason, a non-relativistic
framework provides the most powerful approach to the calculation of the
characteristics of this sort of bound states. In particular, this technique allows one to
evaluate the higher order corrections to the Deser-type formulae systematically. In Refs. \cite{Labelle:1998gh,Eiras:2000rh,Gasser:2001un,Hammer:1999up,Gasser:1999vf} the lifetime of pionium was
evaluated at next-to-leading order in the isospin breaking parameters
$\alpha\simeq1/137$ and $(m_u-m_d)^2$. The non-relativistic framework was
further applied to the ground-state of pionic hydrogen
\cite{Lyubovitskij:2000kk,Gasser:2002am,Zemp} and very recently to the energy-levels and decay widths of kaonic hydrogen
\cite{Meissner:2004jr}.

Within the non-relativistic effective field theory approach we evaluated the S-wave decay widths and strong energy shifts
  of pionium and the $\pi^\pm K^\mp$ atom at next-to-leading order in isospin
  symmetry breaking \cite{Schweizer:2004ir,Schweizer:2004qe}. The width of the
  $\pi^\pm K^\mp$ atom is related to the isospin odd $\pi K$ scattering
  lengths $a_0^-$, while the strong energy shift is proportional to the sum of isospin even and odd scattering lengths
$a_0^++a_0^-$. The values for $a_0^+$ and $a_0^-$, used in the numerical
evaluation of the widths and strong energy shifts, stem from the recent
analysis of $\pi K$ scattering from Roy and Steiner type equations
\cite{Buettiker:2003pp}. Within
chiral perturbation theory (ChPT) \cite{Gasser:1984gg}, the $\pi K$ scattering lengths have been worked out at one--loop accuracy
\cite{Bernard:1990kx,Roessl:1999iu,Nehme:2001wa}, and very recently even
the chiral expansion of the $\pi K$ scattering
amplitude at next-to-next-to-leading order became available \cite{Bijnens:2004bu}.
Particularly interesting is that the isospin even scattering lengths $a_0^+$
depends on the low--energy constant $L_6^r$
\cite{Gasser:1984gg}, and this coupling is related to the flavour dependence
of the quark condensate \cite{Descotes-Genon:1999uh}.

\section{Non-relativistic Lagrangian}
\label{section: Lagr}
 In what follows, we
  concentrate on the spectrum and decays of the $\pi^\pm K^\mp$ atom. The non-relativistic Lagrangian is exclusively determined by symmetries, which are
rotational invariance, parity and time reversal. It provides a systematic
expansion in powers of the isospin breaking parameter $\delta$. What concerns
the $\pi^- K^+$ atom, we count both
$\alpha$ and $m_u-m_d$ as order $\delta$. The different power counting for the
$\pi^+\pi^-$ and $\pi^- K^+$ atoms are due to the fact that in QCD,
the chiral expansion of the pion mass difference $\Delta_\pi=\Mpic^2-\Mpin^2$
is of second order in $m_u-m_d$, while the kaon mass difference
$\Delta_K=\MKc^2-\MKn^2$ starts at first order in $m_u-m_d$. In the sector
with one or two mesons, the non-relativistic $\pi K$ Lagrangian is $\Lagr_{\sss \rm NR} = \Lagr_1+\Lagr_2$. The first term contains the one-pion and one-kaon sectors,
\begin{eqnarray}
  \Lagr_1 &=& \frac{1}{2}(\mathbf{E}^2-\mathbf{B}^2)+h_0^\dagger \Big( i \d_t
  -M_{h^0}+\frac{\Delta}{2M_{h^0}}+\frac{\Delta^2}{8M_{h^0}^3}+\cdots\Big)h_0\nn\\
&&+ \sum_{\pm}h_\pm^\dagger \Big( i D_t
  -M_{h^+}+\frac{\mathbf{D}^2}{2M_{h^+}}+\frac{\mathbf{D}^4}{8M_{h^+}^3}+\cdots\Big)h_\pm,
\label{freeLagr}
\end{eqnarray}
where $\mathbf{E} = -\nabla A_0-\dot{\mathbf{A}}$, $\mathbf{B}= \nabla\times
\mathbf{A}$ and the quantity $h=\pi,K$ stands for the
non-relativistic pion and kaon fields. We work in the Coulomb gauge and eliminate the $A^0$ component of the photon
field by the use of the equations of motion. The covariant derivatives are
given by $D_t h_\pm = \d_t h_\pm \mp i e A_0 h_\pm$ and $\mathbf{D}h_\pm = \nabla
  h_\pm \pm i e \mathbf{A}h_\pm$, where $e$ denotes the electromagnetic coupling. What concerns the one-pion-one-kaon sector, we only list
  the terms needed to evaluate the decay width and the energy shift of the
  $\pi^-K^+$ atom at next-to-leading order in isospin symmetry,
\begin{equation}
  \Lagr_2= C_1'\pim^\dagger\Kp^\dagger\pim\Kp +C_2\left(
    \pim^\dagger\Kp^\dagger\pin\Kn+\textrm{h.c}\right)+
    C_3\pin^\dagger\Kn^\dagger\pin\Kn+\cdots
\label{Lagr}
\end{equation}
The coupling constant $C_1'$ contains contributions coming from the
electromagnetic form factors of the pion and kaon,
\begin{equation}
  C_1' = C_1- e^2\lambda,\quad \lambda =\frac{1}{6}\left( \langle
  r^2_{\pi^+} \rangle+\langle r^2_{\sss K^+} \rangle\right),
\label{charge radii}
\end{equation}
where $\langle
  r^2_{\pi^+} \rangle$ and $\langle r^2_{\sss K^+} \rangle$ denote the charge radii of the charged pion
and kaon, respectively. The low energy constants $C_1,\dots,C_3$ are
  determined through matching to the relativistic theory. For illustration we
  give the result of the matching in the isospin symmetry limit:
\begin{eqnarray}
  C_1 = \frac{2\pi}{\mu_+}\left(a_0^++a_0^-\right),\quad
  C_2 = -\frac{2\sqrt{2}\pi}{\mu_+}a_0^-,\quad
  C_3 = \frac{2\pi}{\mu_+}a_0^+,
\label{CI}
\end{eqnarray}
where the S-wave scattering lengths\footnote{$a_0^+$ and $a_0^-$ are normalized as in
  Ref. \cite{Bernard:1990kx}.} $a_0^+=1/3(a_0^{1/2}+2a_0^{3/2})$ and $a_0^-=1/3(a^{1/2}_0-a^{3/2}_0)$ are defined in QCD, at $m_u
  = m_d$ and $M_\pi \doteq\Mpic$, $M_{\sss K}\doteq\MKc$.

\section{Decay width of the $\pi K$ atom}
To evaluate the energy shift and decay width of the $\pi^- K^+$ atom at
next-to-leading order in isospin symmetry breaking, we make use of
resolvents. For a detailed discussion of the technique, we refer to
Ref. \cite{Schweizer:2004qe}. Here, we simply list the results. The decay width
of the ground-state reads at order $\delta^{9/2}$, in terms
of the relativistic $\pi^-K^+\rightarrow\pi^0K^0$ threshold amplitude \cite{Schweizer:2004ir,Schweizer:2004qe},
\begin{equation}
  \Gamma =
  8\alpha^3\mu_+^2p^*\mathcal{A}^2\left(1+K\right),\quad
  \mathcal{A} = -\frac{1}{8\sqrt{2}\pi}\frac{1}{\Mpic+\MKc}{\rm
  Re}\,A^{00;\pm}_{\rm thr}+\order{(\delta)},
\label{width}
\end{equation}
where $\mu_+$ denotes the reduced mass of the $\pi
K$ atom and
\begin{eqnarray}
  K &=& \frac{\Mpic
  \Delta_K+\MKc\Delta_\pi}{\Mpic+\MKc}(a^+_0)^2-4\alpha\mu_+(a^+_0+a^-_0)\left[{\rm
  ln}\alpha-1\right]+\order{(\delta)}.
\end{eqnarray}
The outgoing relative 3-momentum
\begin{equation}
  p^* = \frac{1}{2 E_1}\lambda\left(E_1^2,\Mpin^2,\MKn^2\right)^{1/2},
\end{equation}
with $\lambda(x,y,z)=x^2+y^2+z^2-2x y-2x z-2y z$, is chosen such that the total
final state energy corresponds to ground-state Coulomb energy
$E_1=\Mpic+\MKc-\alpha^2\mu_+/2$. The quantity ${\rm Re}\,A^{00;\pm}_{\rm thr}$ is calculated as follows. One
evaluates the relativistic $\pi^-K^+\rightarrow\pi^0K^0$ amplitude at order
$\delta$ near
threshold and removes the divergent Coulomb phase. The real part contains
singularities $\sim 1/|\mathbf{p}|$ and $\sim{\rm ln}|\mathbf{p}|/\mu_+$. The
constant term in this expansion corresponds to ${\rm Re}\,A^{00;\pm}_{\rm
  thr}$. The normalization is chosen such that 
\begin{equation}
  \mathcal{A}=a^-_0+\epsilon.
\label{eq: A}
\end{equation}
The isospin breaking corrections $\epsilon$ are calculable within ChPT and have been
evaluated at $\Order{(p^4,e^2p^2)}$ in Refs. \cite{Kubis:2001ij}. For
the numerical analysis, we use the values for the $\pi K$ scattering lengths from the recent analysis of data and Roy-Steiner equations \cite{Buettiker:2003pp}, $a_0^+=(0.045\pm
0.012)\Mpic^{-1}$ and $a_0^-=(0.090\pm 0.005)\Mpic^{-1}$. At next-to-leading
order in isospin symmetry breaking the prediction for the lifetime of the $\pi^\pm
K^\mp$ atom in its ground-state amounts to \cite{Schweizer:2004qe}
 \begin{equation}
    \tau \doteq \Gamma^{-1} = (3.7 \pm 0.4)\cdot 10^{-15}{\rm s}.
  \end{equation}
The bulk part in the uncertainty is due to the uncertainties in the $\pi K$ scattering lengths.   
\section{Energy splittings}
\begin{table}[h]
\tbl{Numerical values for the energy shift of the $\pi^\pm K^\mp$ atom.}{
\begin{tabular}{@{}cccc@{}} \toprule
$\pi^\pm K^\mp$ atom&$\Delta E^{\rm em}_{nl}$[eV]&$\Delta E^{\rm vac}_{nl}$[eV]&
$\Delta E^{\rm h}_{nl}$[eV]\\\colrule
\rule{0mm}{3.5mm}
$n$=1, $l$=0&$-0.095$&$-2.56$&$-9.0\pm1.1$\\
$n$=2, $l$=0&$-0.019$&$-0.29$&$-1.1\pm0.1$\\
$n$=2, $l$=1&$-0.006$&$-0.02$&\\\hline
\end{tabular}}
\label{table: numericspi}
\end{table}
The energy-level splittings of the $\pi^\pm K^\mp$ atom are induced by both
electromagnetic and strong interactions. At order $\delta^3$, the energy shift contributions are exclusively due to strong interactions,
while at order $\delta^4$, both electromagnetic and strong interactions contribute. It is both conventional and convenient to split the energy shifts
into a strong and an electromagnetic part, according to\footnote{Note that this splitting cannot be understood literally, i.e. there are
contributions from strong interactions to $\Delta E^{\rm em}_{n0}$.}
\begin{equation}
  \Delta E_{nl} = \Delta E^{\rm h}_{nl}+\Delta E^{\rm em}_{nl},
\label{DeltaEtot}
 \end{equation}
where $n$ stands for the principal quantum number and $l$ for the angular
 momentum.
 The electromagnetic energy shift contains both pure QED corrections as well as finite size effects due to the
charge radii of the pion and kaon, see Eq. (\ref{charge radii}). The pure electromagnetic corrections have
been evaluated a long time ago by Nandy \cite{Nandy:rj} and the finite size
 effects are given in \cite{Schweizer:2004qe}. For the first two states
 the numerical values for $\Delta E^{\rm em}_{nl}$ are listed in Table \ref{table: numericspi}.
 
At order $\delta^4$, the strong S-wave energy shift yields\cite{Schweizer:2004ir,Schweizer:2004qe},
\begin{equation}
  \Delta E^{\rm h}_{n0} =
  -\frac{2\alpha^3\mu_+^2}{n^3}\mathcal{A}'\left(1+K_n'\right),\quad
\mathcal{A}'=\frac{1}{8\pi(\Mpic+\MKc)}{\rm Re}\,A^{\pm;\pm}_{{\rm
  thr}}+\order{(\delta)},
\label{DeltaE}
\end{equation}
where
\begin{equation}
K_n' = -2\alpha\mu_+(a_0^++a_0^-)\left[\psi(n)-\psi(1)-\frac{1}{n}+{\rm
  ln}\frac{\alpha}{n}\right]+\order{(\delta)},
\end{equation}
and $\psi(n)=\Gamma'(n)/\Gamma(n)$. The quantity ${\rm
  Re}\,A^{\pm;\pm}_{\rm thr}$ is determined by the constant term occurring in
the threshold expansion of the relativistic $\pi^- K^+\rightarrow \pi^- K^+$
amplitude. In the isospin limit, the normalized relativistic amplitude
\begin{equation}
  \mathcal{A}'=a^+_0+a^-_0+\epsilon',
\label{eq: Ap}
\end{equation}
 reduces to the sum of the isospin even and odd scattering lengths. The corrections
$\epsilon'$ have been evaluated within ChPT at $\Order{(p^4,e^2p^2)}$ in Refs. \cite{Nehme:2001wa}. The result for $\Delta E^{\rm h}_{10}$ in Eq. (\ref{DeltaE}) agrees with the one obtained for the strong energy shift of the ground-state in pionic hydrogen
\cite{Lyubovitskij:2000kk}, if we replace $\mu_+$ with the reduced
mass of the $\pi^- p$ atom and ${\rm Re}\,A^{\pm;\pm}_{{\rm
  thr}}$ with the constant term in the threshold expansion for the real
part of the truncated $\pi^- p\rightarrow \pi^- p$ amplitude. The numerical
values for the energy splittings of the first two states are listed in
Table \ref{table: numericspi}, where again the values for $a_0^+$ and
$a_0^-$ are taken from Ref. \cite{Buettiker:2003pp}.

What remains to be added are the contributions coming from the
electron vacuum polarization. The calculation of these corrections within a
non-relativistic Lagrangian approach has been performed in
Ref. \cite{Eiras:2000rh}. In our framework, the contributions due to vacuum
polarization arise formally at higher
order in $\alpha$. However, they are amplified by powers of
the coefficient $\mu_+/m_e$, where $m_e$ denotes the electron mass. For the
first two energy-level shifts, the vacuum polarization contributions $\Delta E^{\rm
  vac}_{nl}$ are listed in Table \ref{table: numericspi}.
\section{Summary}
We provided the formulae for the energy shift and the decay width of the
$\pi^\pm K^\mp$ atom at next-to-leading order in isospin
symmetry breaking. At leading and next-to leading order the $\pi^- K^+$ atom
decays into $\pi^0 K^0$ exclusively. Aside from a
kinematical factor - the relativistic outgoing 3-momentum of the bound system
- the width can be expanded in powers of $\alpha$ and
$m_u-m_d$. The decay width is related to $\pi^- K^+\rightarrow \pi^0 K^0$
threshold amplitude, which is determined in the isospin limit by the isospin
odd scattering length $a^-_0$. The isospin breaking
contributions to this amplitude have been evaluated at
$\Order{(e^2p^2,p^4)}$\cite{Kubis:2001ij} in the framework of ChPT. By invoking ChPT, the result for the decay width
 of the $\pi^- K^+$ atom in its ground-state may be expressed in terms of the
 isospin odd scattering length $a_0^-$, and isospin symmetry breaking
 corrections of order $\alpha$ and $m_u-m_d$, see Eq. (\ref{width}).

It is both conventional and convenient to split the energy shifts into an
electromagnetic and a strong part, according to Eq. (\ref{DeltaEtot}). 
The electromagnetic part contains both pure QED contributions as well as finite
size effects due to the charge radii of the pion and kaon. 
The strong energy shift of the $\pi^- K^+$ atom is
proportional to the one-particle irreducible $\pi^- K^+\rightarrow\pi^- K^+$
scattering amplitude at threshold. In the isospin symmetry limit, the elastic threshold amplitude reduces to the sum of isospin even and odd
scattering lengths $a_0^++a_0^-$. The isospin breaking
contributions to the amplitude have been evaluated at
$\Order{(e^2p^2,p^4)}$\cite{Nehme:2001wa} in the framework of ChPT. The result in
  Eq. (\ref{DeltaE}) displays the S-wave energy shift in terms of the sum $a_0^++a_0^-$, and corrections of order $\alpha$ and
  $m_u-m_d$. 

Our prediction for the lifetime of the $\pi^\pm K^\mp$ atom in its ground-state
yields: $\tau = (3.7 \pm 0.4)\cdot 10^{-15}{\rm s}$, while the numerical
values for the energy shifts are listed in Table
\ref{table: numericspi}. To confront these predictions with data presents a
challenge for future hadronic atom experiments.

\section*{Acknowledgments}
  It is a pleasure to thank J. Gasser, A. Rusetsky, H. Sazdjian and
  J. Schacher for many helpful discussions. This work was supported in part by the Swiss National Science Foundation and by
 RTN, BBW-Contract N0.~01.0357 and EC-Contract HPRN--CT2002--00311 (EURIDICE).

\end{document}